\documentclass[sigconf]{acmart}
\usepackage{bookmark}
\usepackage{hyperref}
\usepackage{amsmath}
\usepackage{amssymb}
\usepackage{booktabs}
\usepackage{tikz}
\usepackage{graphicx}
\usepackage{makecell}
\usepackage{pifont}
\usepackage{subfig}
\usepackage{tabularx}
\usepackage{titlesec}
\usepackage{url}
\usepackage{balance}
\usepackage{xcolor}
\usepackage[normalem]{ulem}

\newcommand{\idea}[1]{\textbf{#1.}}

\newcolumntype{L}{>{\raggedright\arraybackslash}X}
\newcolumntype{C}{>{\centering\arraybackslash}X}
\newcolumntype{R}{>{\raggedleft\arraybackslash}X}

\newcommand{\cmark}{\ding{51}}
\newcommand{\xmark}{\ding{55}}

\usepackage{tikz}
\newcommand{\ballnumber}[1]{\tikz[baseline=(myanchor.base)] \node[circle,fill=.,inner sep=1pt] (myanchor) {\color{-.}\bfseries\footnotesize #1};}

\renewcommand\footnotetextcopyrightpermission[1]{}

\begin{document}
\pagestyle{plain}
\title{CoinPolice: Detecting Hidden Cryptojacking Attacks with Neural Networks}

\author{Ivan Petrov}
\affiliation{Lomonosov Moscow State University}
\email{ipetrov@cs.msu.ru}

\author{Luca Invernizzi}
\affiliation{Google}
\email{invernizzi@google.com}

\author{Elie Bursztein}
\affiliation{Google}
\email{elieb@google.com}

\begin{abstract}

Traffic monetization is a crucial component of running most for-profit online businesses.
One of its latest incarnations is cryptocurrency mining, where a website instructs the visitor's browser to participate in building a cryptocurrency ledger  in exchange for a small reward in the same currency.

In its essence, this practice trades the user's electric bill, or battery level, for cryptocurrency.
With  user consent, this exchange can be a legitimate funding source -- for example, UNICEF is collecting donations in this fashion on \url{thehopepage.org}.
Regrettably, this practice also easily lends itself to abuse: in this form, called \emph{cryptojacking}, attacks surreptitiously mine in the users browser, and profits are collected either by website owners or by hackers that planted the mining script into a vulnerable page.

Understandably, users frown upon this practice and have sought to mitigate it by installing blacklist-based browser extensions (the top 3 for Chrome total over one million installs), whereas researchers have devised more robust methods to detect it.
In turn, cryptojackers have been bettering their evasion techniques, incorporating in their toolkits domain fluxing, content obfuscation, the use of WebAssembly, and throttling.
The latter, in particular, grew from being a niche feature, adopted by only one in ten sites in 2018~\cite{konoth2018minesweeper}, to become commonplace in 2019, reaching an adoption ratio of 58\%.
Whereas most state-of-the-art defenses address multiple of these evasion techniques, none is resistant against all.

In this paper, we offer a novel detection method, CoinPolice, that is robust against all of the aforementioned evasion techniques.
CoinPolice flips throttling against cryptojackers, artificially varying the browser's CPU power to observe the presence of throttling.
Based on a deep neural network classifier, CoinPolice can detect 97.87\% of hidden miners with a low false-positive rate (0.74\%).
We compare CoinPolice performance with the current state of the art and show our approach outperforms it when detecting aggressively throttled miners.

Finally, we deploy Coinpolice to perform the largest-scale cryptoming investigation to date, identifying 6700 sites that monetize traffic in this fashion.

 \end{abstract}

\maketitle

\section{Introduction}
Cryptocurrencies are digital assets that are exchanged primarily without the involvement of trusted parties, such as banks or government institutions. Their ownership, instead, is recorded in distributed ledgers, maintained by mutually-distrustful parties, called \emph{miners}.
These ledgers are sprinkled with easy-to-verify but hard-to-generate cryptographic proofs, intended to make the generation of forged ledgers extremely computationally expensive, while keeping the ledger verification cost low. This computational asymmetry is where the trust among miners is rooted.

Miners incur ongoing power and hardware costs to maintain these ledgers, so they are compensated with a cryptocurrency reward each time a cryptographic proof is successfully generated. To reduce these costs and increase profits, miners have focused on perfecting efficiency, switching general purpose hardware such as CPUs/GPUs with dedicated integrated circuits (ASICs), which can outperform CPUs by a factor of up to 10 million.
The adoption of ASICs and the success of the popular cryptocurrency Bitcoin has lead to an explosion of the computational power behind cryptocurrencies: the Bitcoin network has seen an increase of 750X~\cite{bitcoincharts} between January 2016 and 2019. As cryptocurrency networks adjust the difficulty of the cryptographic proofs according to the available computational power, miners without dedicated hardware have seen their profits turn into losses, and have been thus priced out of the market.

With the explicit intent to make mining more egalitarian and increase miner diversity,
researchers have created cryptocurrencies based on \emph{ASIC-resistant hash algorithms}, with the CryptoNight~\cite{van2013cryptonote} algorithm leading the pack in terms of adoption.
Cryptonight is designed to give a competitive advantage to CPUs, thus leveling the field.

The surge in popularity of Cryptonight-based cryptocurrencies, such as Monero, in conjunction with the near-native computational efficiency of WebAssembly, a web scripting language, has opened the door to in-browser cryptocurrency mining.
This practice has been poised as an alternative to ads, as this is effectively a traffic monetization mechanism, turning visits into (crypto) cash. Riding on this new technology, a variety of online services has been launched to offer easy API to ``Monetize Your Business With Your Users' CPU Power'' (CoinHive's motto). These services have been used to collect charity donations with  user consent (e.g, Unicef), or as an alternative to ads (e.g., by Salon.com and ThePirateBay).
Recently, the most popular of these services, CoinHive, has ceased operations, due to a change in the underlying cryptocurrency.

Regrettably, this monetization mechanism is also ripe for abuse, as it allows website owners or hackers to effectively spend the unaware visitor's power, and the utility bill that comes with it, to earn cryptocurrency.
This type of attack is called \emph{cryptojacking} (also known as \emph{drive-by-mining}), and previous studies \cite{eskandari2018first,hong2018you} estimate its yearly revenue to be at least 20M USD.

As an example, we quantify the harm that cryptojacking inflicts to a mobile user in Figure~\ref{figure:battery}. A non-throttled miner running on the background can shorten the battery life of a latest-generation high-end mobile device by over 75\%\footnote{See Section~\ref{sec:battery} for detail on this measurement.}.

Tech-savvy users have sought to mitigate cryptojacking by installing browser extensions, such as NoCoin \cite{nocoin}, that blacklist the URLs whence the mining scripts are being loaded. As these solutions are fraught with the usual issues that blacklist face (e.g., their reactive nature, high upkeep burden, and easy bypassability), researchers have proposed more sophisticated detection mechanisms ranging from static analysis to behavior classification. To protect their income, cryptojackers have also been bettering their evasions toolkit, borrowing techniques from the malware community such as domain fluxing and content obfuscation. To remain unnoticed, cryptojackers have also turned to throttling their mining speed: whereas this practice was relatively uncommon at the start of 2018, with only one in ten~\cite{konoth2018minesweeper} miners using it, our 2019 sample of 200 miners found that throttling has now been adopted  by the lion share of miners\footnote{$H_0$=50\% of mining sites, significance level=0.05, observed proportion=58\%, sample size=200, p=0.023.}. Moreover, major mining service providers have included throttling in their "quick start" example code, boosting its adoption. Unfortunately, we have found these evasion techniques to be effective in bypassing the state-of-the-art detection mechanisms (see Figure~\ref{figure:throttling}).

In this paper, we present CoinPolice, a code-agnostic cryptojacking detection system. With CoinPolice, we look for in-browser execution patterns, either of JavaScript (\emph{JS}) or WebAssembly (\emph{WASM}) code, that are repetitive in nature and that actively adjust to the computational power of the execution environment, so to keep their CPU time within a target fraction of the available computing power. To the best of our knowledge, this is a novel approach to this problem. CoinPolice executes web pages in the Chrome browser, and leverages its Developer Tools API to collect execution traces and to artificially slow down the execution speed of the site. Having collected multiple traces at various execution speeds, it then feeds this data into a deep neural network that will pinpoint any mining scripts. We test CoinPolice on a dataset comprising of $\sim$47k samples to show that it can detect cryptojacking with 97.8\% accuracy and
CoinPolice can detect malicious cryptocurrency mining on web pages with 97.8\% true-positive rate and 0.8\% false-positive rate.

Finally, we deploy CoinPolice to perform the largest-scale cryptomining investigation to date, identifying 6700 sites with functioning hidden miners.

In summary, the main contributions of this paper are:
\begin{itemize}
  \item We show that cryptojackers are relying more on evasion techniques, such as throttling, and this is rendering state-of-the-art defenses less effective
  \item We propose CoinPolice, a novel detection approach based on the active variation of the execution environment of cryptominers, which is capable of detecting heavily-throttled miners.
  \item We perform the largest-scale study of cryptomining prevalence to date, to identify 6700 sites with active miners.
  \item We measure the recovery rate of cryptojacking sites that were  affected by the ceasing of operation of the largest cryptomining-as-a-service provider, CoinHive.
\end{itemize}
 \section{Background}
In this section, we provide a succinct overview of key cryptocurrency and
cryptojacking concepts, sd to make this paper more accessible to readers
unfamiliar with these topics.  Feel free to skip.
\subsection{Cryptocurrency Mining}

Cryptocurrencies are digital assets that are exchanged among mutually-distrustful parties in a peer-to-peer network. They are commonly being used to pay for goods and services, or as investment vehicles.
In place of a central authority, such as a bank, keeping track of the ownership of these assets, cryptocurrencies use a distributed ledger, called the \emph{blockchain}. As the name implies, this is an ever-growing sequence of \emph{blocks} which bundle the transactions happening on the network. Blocks are interlinked, as each block contains a hash of the previous one. This chaining makes it extremely computationally expensive to falsify past transactions, as this would require to generate a hash collision.
Additionally, the blockchain is protected against denial of service attacks and forgeries by \emph{proofs of work} (PoW) embedded in each block. These are asymmetrical cryptographic challenges that are hard to compute, so to slow down an attacker aiming to create alternate chains, but easy to verify, so to allow the swift detection of forgeries.

A peer-to-peer network of hosts, called \emph{miners}, maintains the blockchain, collecting transactions and crystallizing them into new blocks that get appended to the blockchain. To do so, these hosts have to compute the expensive PoW in a process called \emph{mining}. When a PoW is computed successfully and a block is generated, the miner responsible is rewarded with some amount of cryptocurrency.
The difficulty of PoWs is dynamic and varies with the number of miners, so to allow the block generation rate constant.

The generation of a valid block is a winner-take-all horse race: the miner that solves the PoW first is rewarded, and the rest get nothing. The reward can be substantial: for Bitcoin, the current reward is 12.5 bitcoins, which equals \$65.025~\cite{bitcoinreward}. As the probability to win this race decreases with the network size, some miners have opted to organize into \emph{mining pools}.
When one of the participants wins the horse race, the pool collects the reward, takes a cut and distributes the rest to the pool participants, according to the number of contributed shares.

\subsection{CPU Mining}
The competitive nature of mining has pushed the miners to seek the bleeding edge of performance. Professional miners have moved from mining on general purpose CPUs to the more adept to batch parallelization GPUs, then opting for specialized programmable hardware (FPGA), to finally produce single-purpose integrated circuits (ASICs) that are capable of mining faster than 10 million CPUs.
This race to performance has increased the PoW difficulty, decreasing the probability of payoffs for miners that remained on CPUs. As the power/hardware cost to mine remain constants, this race has priced CPU miners out of the market.

ASIC mining has raised the cost to enter the mining market, reducing the number of agents operating in the market. As this poses a threat to the decentralized nature of cryptocurrencies, researchers have devised PoW algorithms that give a competitive advantage to CPU mining, allowing these smaller miners to be profitable again. One of the most popular is \emph{CryptoNight}~\cite{van2013cryptonote}. This algorithm requires large amounts of random-access memory reads from a two-megabyte array, called a \emph{scratchpad}.
This array fits in the extremely-fast L3 cache of modern CPUs, but it poses a considerable challenge to ASIC mining, as ASICs are ill-suited to hold large arrays. GPU mining is also impacted, as L3 cache is faster than the typical memory used on GPUs, GDDR5.

\subsection{In-Browser Mining}
In-browser mining is the idea of mining cryptocurrencies directly in web pages, through JS or WASM code.
It was first introduced in 2011 to mine Bitcoin~\cite{jsminer} but, as this cryptocurrency uses an ASIC-friendly PoW algorithm, it quickly became pointless, as the generated revenue approached zero.
In-browser mining bounced back from obsolescence and into profitability due to two technologies: the introduction of ASIC-resistant currencies, such as \emph{Monero}~\cite{monero} which uses CryptoNight under its hood, and the proliferation of WASM (announced in 2015), which allows for more efficient in-browser scripting.

The first to introduce in-browser mining as-a-service was \emph{Coinhive}~\cite{coinhive}. This service consisted of a mining pool coupled with an easy-to-deploy mining script that just needed to be copied into a web page source code.
Soon, new competitors entered this new market, including Crypto-Loot \cite{cryptoloot}, and CoinImp \cite{coinimp}, poising
in-browser mining as-a-service as an traffic-monetization alternative to ads.

\subsubsection{Impact on battery life}\label{sec:battery}
The harm inflicted on cryptojacking victims depends on the  hardware used in the attack, and the context. For desktops and plugged-in laptops/mobile devices, the harm consists mainly of an increased electric bill, lower performance, and heat. This attack is more harmful to users with mobile devices running on battery power, as it shortens the battery life span and  eventually causes the device to shut down.

\begin{figure}[ht]
    \centering
    \includegraphics[width=1.0\linewidth]{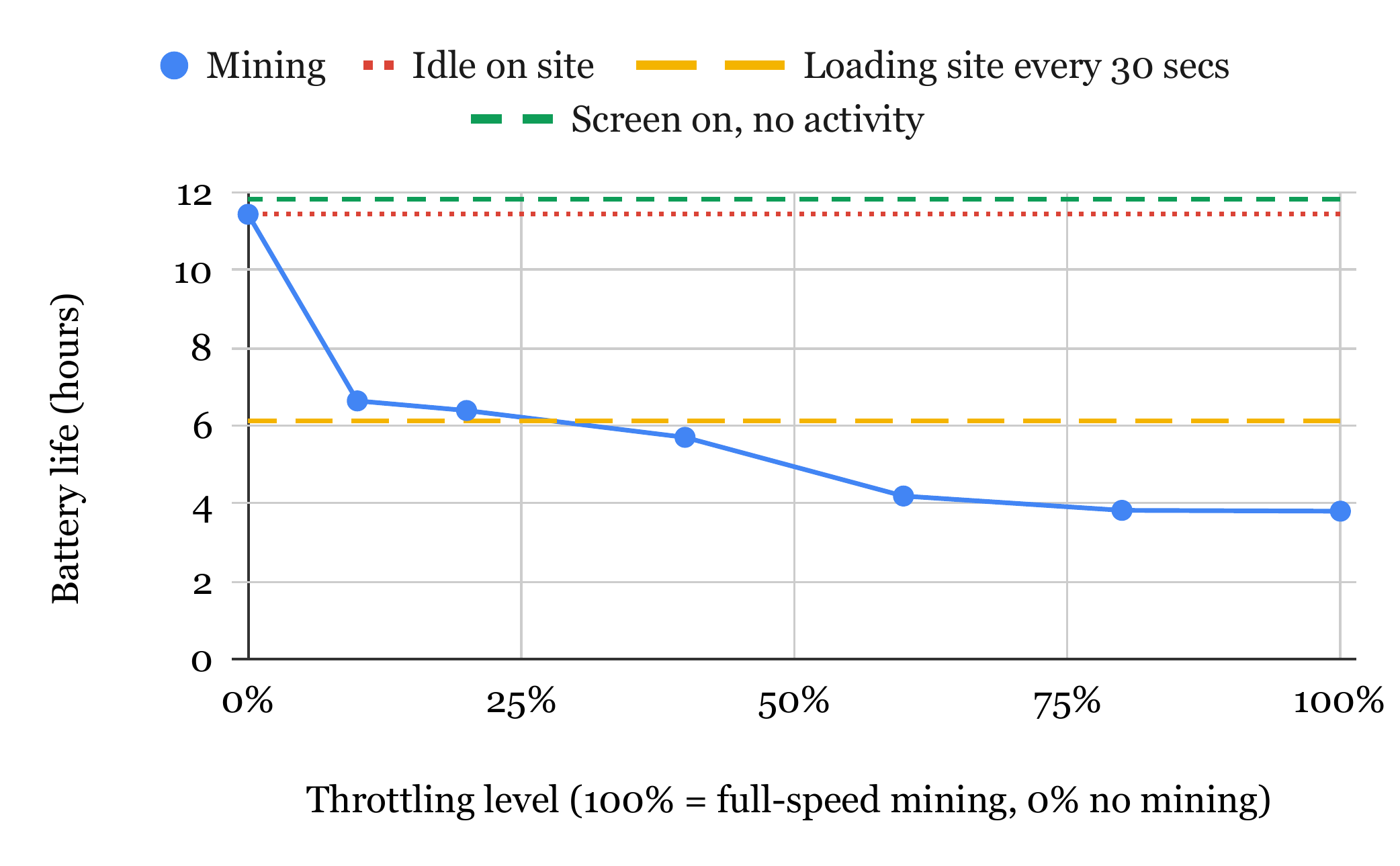}
    \caption{Impact on battery life of a cryptojacking attack.}
    \label{figure:battery}
\end{figure}

We quantify this latter effect in Figure~\ref{figure:battery}. For this experiment, we run a high-end phone (released in 2018)  in a variety of configurations and measure the power drawn from the battery, so to estimate the battery life span. We ran a total of nine configurations: \ballnumber{1} device is idle with its screen turned on, \ballnumber{2} same, but the  Chrome browser is open on a site (\url{thehopepage.org}) with no mining occurring, \ballnumber{2} same, but the browser is reloading the page every 30 seconds, and \ballnumber{3-9} the opened site is  mining at different throttling levels\footnote{Note that some throttling levels cannot be achieved with the controls given in \url{thehopepage.org}. We injected t code into the page to set these levels.}.
We run each configuration for thirty minutes, measuring the battery level through ADB commands at the beginning and end of the run. The results show that a miner running at full speed can severely reduce the battery life span to less than four hours on a full charge, but even a 10\% throttling level can reduce the battery life span significantly.

Our empirical results show, then, that heavily-throttled cryptojacking causes almost the same level of mobile users' harm as full-speed mining. This is because the mobile device is never idle when mining, and as such it cannot enter its most power-efficient states.

We note  several confounding variables  affect these measurements, such as background activity, and the device temperature. In particular, the latter increases swiftly under heavy mining load: for example, it rose 11.3$^\circ C$ (from 26.3$^\circ C$) after ten minutes of full-speed mining. Eventually, the CPU might slow down its execution speed to avoid overheating, depending on ambient temperature and the device cooling capability.

 \section{Threat model}\label{section:threatmodel}
\idea{Intent} The intent of the adversary we are facing is to collect cryptocurrency payouts through cryptojacking. The non-abusive form of in-browser mining, where users are informed of the activity and consent to it, is out of scope. In our evaluation (Section~\ref{section:evaluation}), we take precautions to make sure we never consent to mine when browsing.

\idea{Delivery vectors} Adversaries can insert mining scripts into websites they own, or into JavaScript/WASM libraries they distribute to third parties. Additionally, they can hack into vulnerable websites and third-party libraries to inject mining scripts into them.

\idea{Evasion techniques}  Adversaries can craft mining scripts in JavaScript, WebAssembly, or both. They can split the mining logic into multiple collaborating scripts, and obfuscate their code. They can execute these scripts in the main browser execution thread for that page, or in separate threads, such as WebWorkers, SharedWebWorkers or ServiceWorkers. They can leverage domain fluxing to evade blacklists. Finally, they can throttle the execution speed of their mining scripts to remain undetected.

\idea{Motivation}  We choose to give our model adversary a vast toolset of evasion techniques to reflect the capabilities of real attackers. Most of these techniques have been observed in the wild by previous research: for example, domain fluxing have been observed in~\cite{konoth2018minesweeper},
the use of WebAssembly in~\cite{wang2018seismic}, and code obfuscation in~\cite{liu2018novel}.

The use of throttling is relatively new in the ecosystem, at it has only been measured by one previous study~\cite{konoth2018minesweeper}, to be in use by only 11.36\% of mining scripts.

As a preliminary measurement of the incidence of throttling in miners in March 2019, we have collected 200 URLs of sites that use the most popular mining service at the time, CoinHive~\cite{coinhive}. We did so by querying a public search engine that indexes website source code, PublicWWW~\cite{publicwww}. Then, we visited each page with Chrome to identify the mining script and extract the throttling logic to identify the throttling level. We have chosen to do this preliminary measurement without leveraging  CoinPolice, so not to be biased by its detection. We observe in  that the lion share of websites have adopted throttling, and 8.5\% are throttling aggressively (that is, less than 25\% of computing power is used to mine). The peak at the 30\% throttling rate is likely caused by this value being the default in CoinHive's quick-start tutorial. We will leverage Coinpolice to expand this analysis to other mining libraries in section~\ref{section:evaluation}.

\begin{figure}[htbp]
    \centering
    \includegraphics[width=1.0\linewidth]{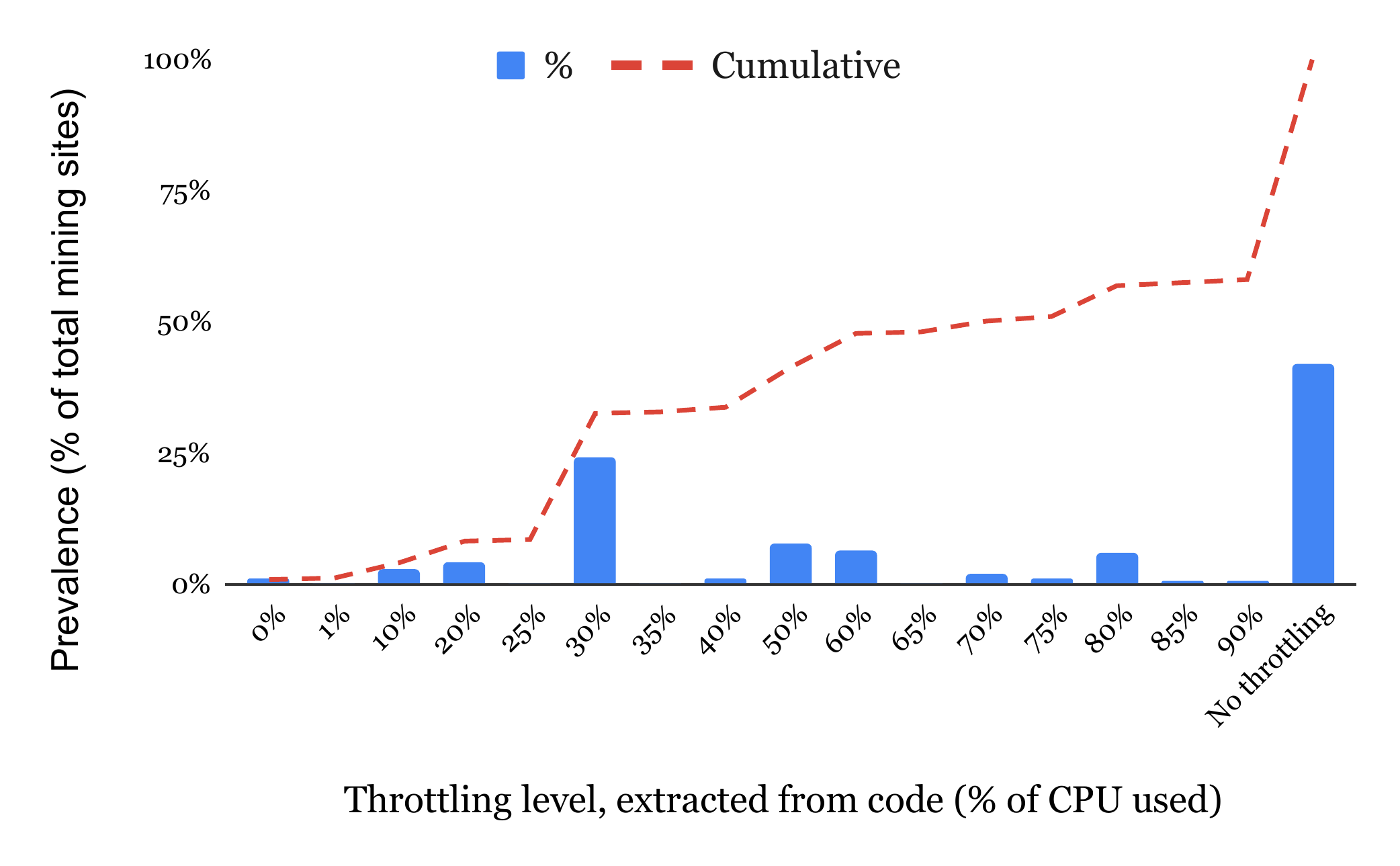}
    \caption{Distribution of throttling levels in a random sample of 200 CoinHive scripts.}
    \label{figure:coinhive_throttling}
\end{figure} \section{Related Work}\label{section:related}

Since the surge of cryptojacking in 2017, researchers have proposed several methods to detect and mitigate it. In this section, we categorize and summarize these proposals, along with their advantages and disadvantages. See Table~\ref{table:relatedwork} for a summary.

\idea{Static detection} The first solutions to combat cryptojacking are based on blacklisting, so to prevent the URLs of mining pools and proxy services from being loaded. These solution are usually offered in the form of browser extensions, such as NoCoin \cite{nocoin}, AntiMiner \cite{antiminer}, MinerBlock \cite{minerblock}. Incidentally, blacklists are the only detection mechanism that has been adopted by a significant number of users, with more that one million total installs for the top three extensions. Blacklist-based approaches, however, are fraught with drawbacks: they can easily be bypassed by domain/URL fluxing, as they are reactive in nature, and because of it they incur an high upkeep burden.
Perhaps because of the popularity of these blacklists, over 80\% of cryptojackers~\cite{konoth2018minesweeper} are using domain/URL fluxing.

These limitations challenge the viability of blacklist-based approaches in the long term, as acknowledged by Chromium developers~\cite{chromiumbug}. In fact, the effectiveness of blacklists is already declining: Hong et al.~\cite{hong2018you} note that blacklists are updated every ten days on average, which is less frequent than the churn rate of cryptojacking domains.

A more advanced static detection approach has been proposed in MineSweeper~\cite{konoth2018minesweeper}, which used statically analyzes WASM instructions to identify loops which have a high number of XOR, shift, and rotate operations. As the authors state, ``in the future, cybercriminals may well decide to sacrifice profits and highly obfuscate their cryptomining Wasm modules to evade detection. In that case, the previous algorithm is not sufficient.'' For this, they propose to monitor CPU-cache events, which we will discuss in the next section. Unfortunately, the authors do not provide the classification performance for MineSweeper.

Finally, Saad et al.~\cite{saad2018end} propose to detect cryptojacking by clustering similar JS code. To do so, they apply the fuzzy C-means clustering algorithm to code-complexity features extracted from a reconstructed control-flow graph. Although the authors claim a 96\% accuracy and 3.3\% false positives, it is unclear how this approach is scalable, as it was tested on only eight cryptojacking scripts, and code obfuscation and WASM were not considered.

\idea{Dynamic hardware-based detection} This class of detection approaches aims to profile the execution of cryptomining code at the CPU level, using the hardware performance counters (HPCs) included in modern CPUs. These counters are CPU registries that keep a tally of internal CPU events, such as the number of executed cycles, cache misses, branch prediction exceptions, so to provide developers with accurate metrics on the performance of their code. As CryptoNight is highly optimized to fit the L3 cache of modern CPUs, the intuition is that its execution will be detectable when analyzing a timeseries of HPC events.

As this approach is not open source, we have implemented this approach to the best of our abilities, so to compare the effectiveness of CoinPolice with it and evaluate it fairly. Our improvements of this approach over the state of the art includes adding a deep-learning based classifier in place of hard-coded rules, and reporting classification performance numbers.

This class of detection approaches has several drawbacks: as multiple processes/threads are scheduled on the same CPU, the HPC values only give an aggregate view of the activities that are happening on the host. This impedes the detection of throttled miners, as their activity can become indistinguishable from a WASM-based game, video player, or CPU-intensive 3d visualization. This approach also poses scalability challenges: the aggregation that HPC impose impedes parallelization on the same CPU/host, and the most prominent cloud providers such as Amazon Web Services and Google Cloud, prevent their clients from accessing HPC values.

\idea{Dynamic browser-based detection} The most active area of research is centered on profiling the execution of cryptojacking scripts and detect their behavior. Our approach, CoinPolice, also falls into this class. Two main categories of approaches have emerged: detecting anomalous timeseries of specific WASM instructions, and profiling the overall browser execution.

The former approach has been followed by SEISMIC~\cite{wang2018seismic}, and CMTracker~\cite{hong2018you}. In the former, WASM code is augmented with inline counters that tally the execution count of five opcodes (\emph{i32.add}, \emph{i32.and}, \emph{i32.shl}, \emph{i32.shr\_u} and \emph{i32.xor}), incurring an execution overhead of 100\%. These features are then fed into a support-vector machine classifier. The authors claim a 98\% accuracy, but it is unclear if this approach is feasible at scale, as it has only been tested on twelve WASM scripts. Such an approach is likely to be vulnerable to obfuscation, or to splitting the mining execution over multiple collaborating JS/WASM scripts.

In CMTracker, the authors select nine JS function names as signatures (such as \texttt{cryptonight\_hash}, or \texttt{crypto}) and compute their total execution time. If a threshold of 10\% is surpassed, the script is classified as mining. Secondly, they identify repeating function call sequences, and classify the script as mining if more that 30\% of the execution time is spent in one of these sequences. Unfortunately, the authors do not report classification  performance values. Such an approach does not consider WASM mining, which accounts for the vast majority of the miners we have encountered. Plus, it can be evaded by a combination of throttling and minimal code obfuscation (or even minification).

The related work more relevant to ours is  OUTGUARD~\cite{outguard}, which computes a series of behavioral features to then classify them through an SVM (support vector machine). These features include the number of web workers, the number of workers executing identical tasks, websocket creation count, specific function names (such as \texttt{CryptonightWasmWrapper.hash}), the number of messages sent to WebWorkers, and the number of MessageLoop events.
The key differences between Outguard and our work stem from our design decision  not to rely on rely on \ballnumber{1} signature matching on the attacker code, which could be obfuscated, or \ballnumber{2} features designed to catch aggressive mining, such as the number of web workers running identical tasks. These choices allow Coinpolice to be agnostic to the implementation details of the mining, thus increasing its robustness. Additionally, Coinpolice's detection is not skewed toward catching aggressive miners so to be able to better detect heavily-throttled miners, which represent a growing trend in this space. OUTGUARD does not provide classification performance data at varying mining throttling levels.

\begin{table*}[htbp]
\caption{Related work}
\begin{center}
  \small
  \begin{tabularx}{\linewidth}{lp{4.2cm}p{3cm}L}
  \toprule
  \textbf{Paper} & \textbf{Features} & \textbf{Classifier} & \textbf{Performance}\\
  \midrule
  \textbf{CMTracker~\cite{hong2018you}} & JS signatures & Hand-crafted rules & Not reported \\
  \textbf{SEISMIC~\cite{wang2018seismic}} & WASM opcode counts & Hand-crafted rules& 98\% accuracy, calculated on 12 samples \\
  \textbf{MineSweeper~\cite{konoth2018minesweeper}} & Human-assisted static analysis  & Hand-crafted rules & Not reported \\
  \textbf{Saad et al.~\cite{saad2018end}} & Clustering code-complexity features&  Fuzzy C-means & 96\% accuracy and 3.3\% false positives, calculated on only 8 samples\\
  \textbf{OUTGUARD~\cite{outguard}} & Signatures and behavioral features of aggressive miners  & Support-Vector Machine &97.9\% TPR and 1.1\% FPR, trained and tested on 30k samples (10\% test split) with no throttiling randomization.\\
  \midrule
  \textbf{Coinpolice (this work)} & Throttling-independent behavior timeseries & Convolutional neural network & 97.8\% TPR and 0.74\% FPR, trained and tested on 47k samples (10\% test split) with \textbf{randomized throttling levels}.\\
  \bottomrule
\end{tabularx}
\label{table:relatedwork}
\end{center}
\end{table*} \section{Detection Algorithms}\label{section:algos}

As the majority of related work either do not provide classifier performance data or a functional open-source implementation, we have chosen to design and implement to the best of our ability several classifiers that replicate the strategies they use. These classifiers are not a perfect reproduction of related work: instead, we have opted to implement similar features and process them through a deep neural network classifier. We do so to push each detection strategy to the limit, and offer a fairer comparison between CoinPolice and other strategies, as deep-learning detectors typically match or outperform more traditional approaches.

Specifically, we describe seven cryptojacking detection classifiers: \ballnumber{1} a baseline classifier that only uses an hard-coded 30\% threshold over CPU usage, \ballnumber{2,3} two classifiers based on HPC counters (respectively using a convolutional and recurrent neural networks), \ballnumber{4,5} two classifiers based on JS/WASM function execution timeseries, and \ballnumber{6,7} two classifier based on the JS/WASM features and throttling-detection features (one being CoinPolice, and a convolutional deep-learning classifier using the same features).

The mapping between classifiers and the features they use is provided in Table~\ref{table:featuremapping}.

\begin{table}[htbp]
  \caption{Features used in each classifier.}

\begin{center}
  \footnotesize
\begin{tabularx}{\linewidth}{lCCC}
  \toprule
  \textbf{Classifier} & \multicolumn{3}{|c}{\textbf{Features}} \\
&  \multicolumn{1}{|L}{Hardware-performance counters} & Recurring code execution & Throttling detection\\
  \midrule

  HPC CNN & \cmark & \xmark &\xmark \\
  HPC RNN & \cmark &\xmark &\xmark \\
  \midrule
  JS CNN &\xmark & \cmark &\xmark \\
  JS RNN &\xmark & \cmark &\xmark \\
  \midrule
  JS CNN &\xmark & \cmark & \cmark \\
  JS RNN (CoinPolice) &\xmark & \cmark & \cmark \\
  \bottomrule
\end{tabularx}
\label{table:featuremapping}
\end{center}
\end{table}

\begin{figure}[ht]
    \centering
    \includegraphics[width=1.0\linewidth]{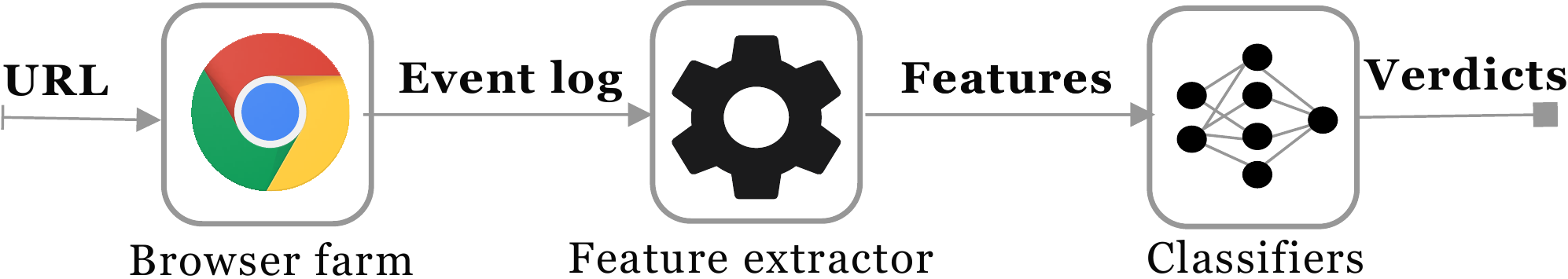}
    \caption{Overall system overview.}
    \label{figure:overview}
\end{figure}

The overall system design is shown in Figure~\ref{figure:overview}: the input of the system is a URL, which we crawl with a browser farm of Chrome browsers.  We obtain a log of events that occurred at the system level and within the browser while the URL was loaded in a tab, and extract the features described in Section~\ref{section:features}. Finally, we feed the features into the classifiers to obtain a verdict from each classifier.

We note that the classifiers \ballnumber{4-7} exclusively use data obtained from the Chrome DevTools API. As such, they could be executed both server side or client side. The HPC-based classifiers, instead, are hardware dependent and susceptible to system noise. As such, a server-side deployment would be preferable, both for the reduced complexity and increased signal-to-noise ratio (see Section \ref{section:hpc}).

As HPC-based classifiers incur in significant deployment challenges (see Section~\ref{section:hpc}) we also chose not to include in our evaluation a classifier that uses all features: its deployment would be challenging, and our results show it would provide no classification benefits over CoinPolice.

\subsection{Features}\label{section:features}
We collect three categories of features: CPU hardware performance counters, features describing cyclical patterns extracted from JS/WASM execution traces,  and features characterizing the changes in the same traces in response to active CPU execution speed slowdowns. We will now describe the concepts underlying these features in detail, and we will discuss their representation in the next subsection.

\subsubsection{Features: Hardware-Performance Counters}\label{section:hpc}

HPCs are counters embedded into modern CPUs to keep a tally of internal events that affect the execution performance. The specific counters are hardware dependent but, generally, they measure the number of executed cycles, cache misses, and branch prediction exceptions. Originally, HPCs were provided to developers to surface execution bottlenecks in their programs. Being implemented in hardware, the overhead incurred by monitoring HPCs is minimal.

Here, we use HPCs to discriminate benign/malicious execution thanks to the key property of a mining algorithm designed to give a competitive advantage to CPU miners, such as  CryptoNight -- its memory hardness.  For example, CryptoNight requires substantial memory reads (524,288 reads per hash)~\cite{van2013cryptonote} from a pseudo-randomly generated array, which is designed to fit on the L3 cache of modern CPUs. We note that this approach is not specific to CryptoNight, but generalizes to any algorithm that uses the speed of CPU caches to give a competitive advantage to CPU miners. In the rest of the paper, we often refer to CryptoNight as it is by far the most common algorithm used to give this advantage.
When a non-obfuscated, non-throttled version of CryptoNight is being executed, these memory reads rarely cause L3 cache misses, which effectively makes CryptoNight-based mining competitive on CPUs.

\begin{figure}[ht]
    \centering
    \hfill
    \begin{minipage}{1\linewidth}
        \subfloat[L3 cache loads.\label{figure:llc_loads}]{\includegraphics[width=0.49\linewidth]{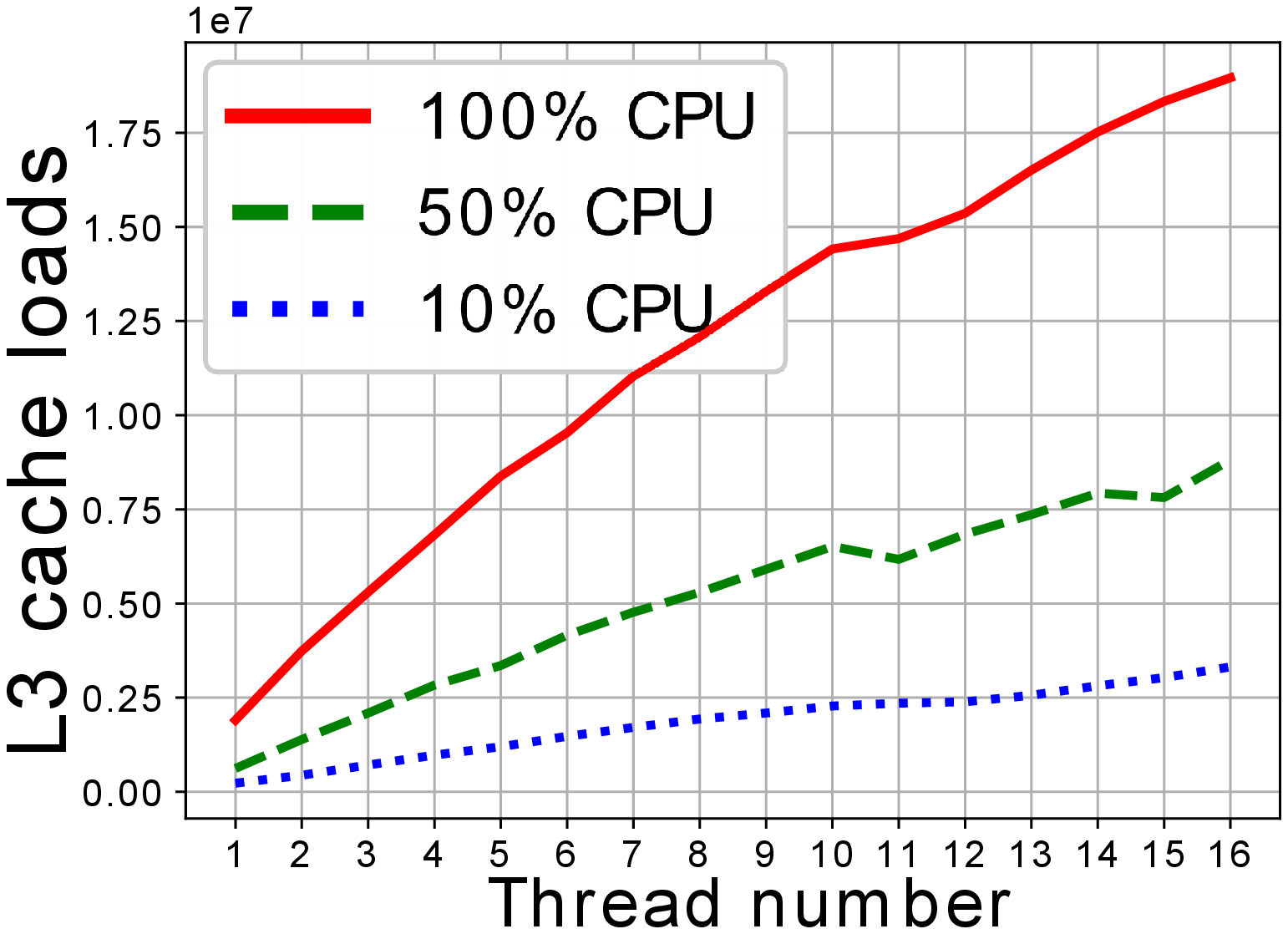}
        }
        \hfill
        \subfloat[L3 cache load misses.\label{figure:llc_load_misses}]{\includegraphics[width=0.49\linewidth]{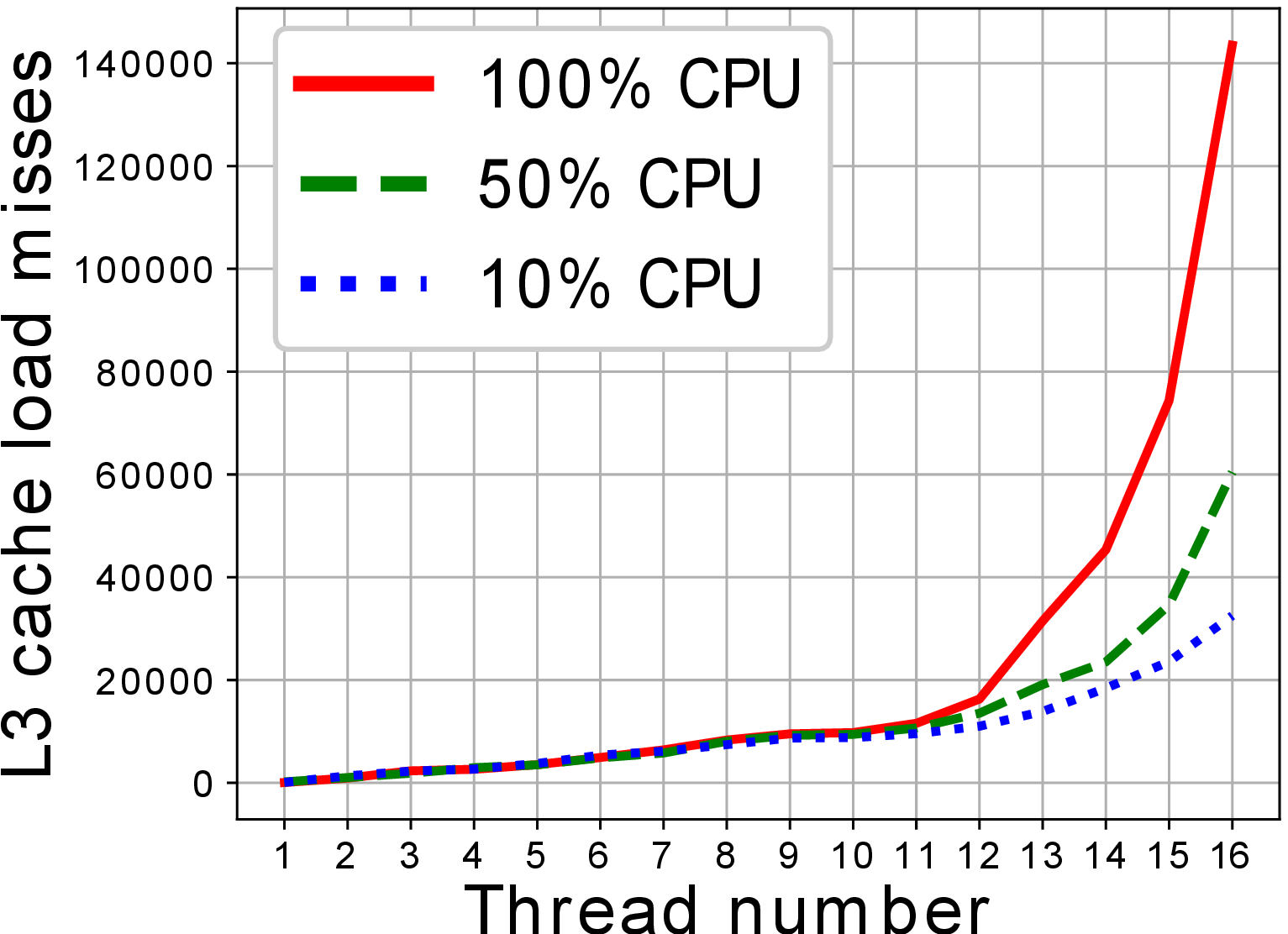}
        }
        \caption{L3 cache loads/misses (server had 35,840 KB L3 cache size).}
        \label{figure:llc_loads_load_misses}
    \end{minipage}
\end{figure}

\idea{Limitations}

HPC-based features are highly susceptible to noise introduced by other processes or threads running on the same host, as they all share the same CPUs, and compete to store values in their caches, causing cache misses. Since users encounter cryptojacking miners while browsing the web, the HPC values will be affected by all open tabs, severely diminishing the discriminative power of this feature. Additionally, cryptominers can intentionally cause additional cache misses to further impede detection, or offer to users compelling content to keep them on the site longer, which also will compete for computational resources. Miners that parallelize their activity with WebWorkers to increase the hash rate will also increase the cache-miss rate (see Figure~\ref{figure:llc_loads_load_misses}).
Finally, HPCs pose practical challenges. They are hardware dependent, so classification performance will vary across CPU manufactures and models, and trained models might not be easily transferable.

\subsubsection{Features: Recurring Code Execution}
\begin{figure}[htbp]
    \centering
    \includegraphics[width=1.0\linewidth]{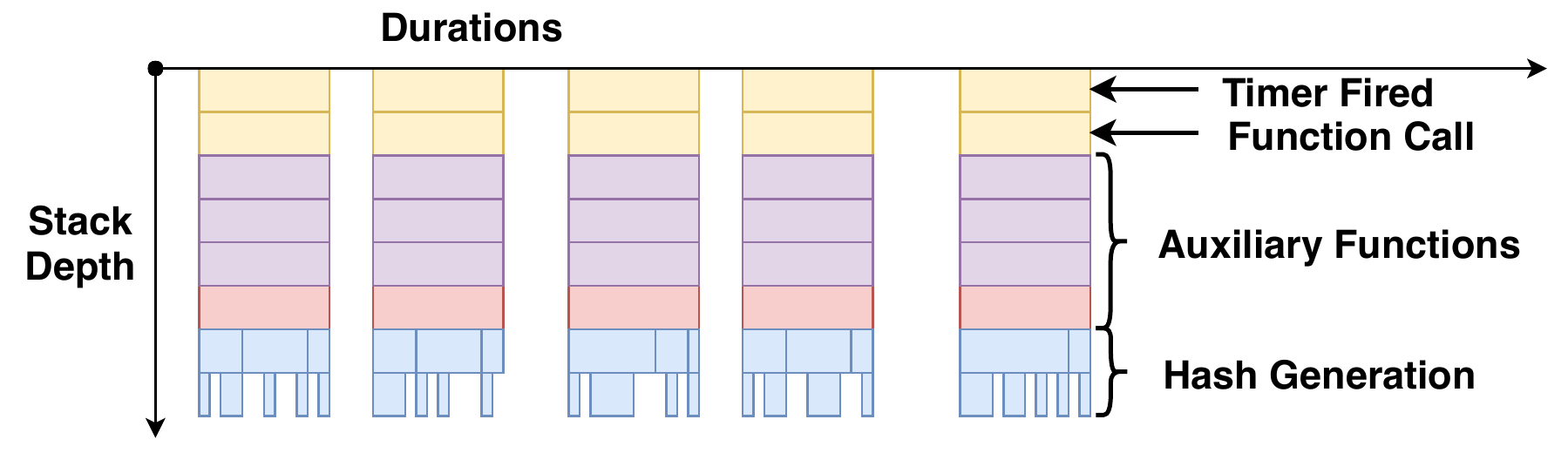}
    \caption{Stack trace timeline for a Coinhive script.}
    \label{figure:stack_trace_timeline}
\end{figure}

Cryptocurrency mining is innately a repetitive process, where hashes are produced over and over.
This repetition is apparent when looking at JS/WASM stack traces timeseries. This is exemplified in Figure~\ref{figure:stack_trace_timeline}, which shows a stack trace timeline of a Coinhive mining script collected via \emph{Chrome DevTools}~\cite{devtools}.
Note that the same set of functions is being executed repeatedly, with similar sequencing and durations. This pattern is relatively uncommon in legitimate websites - according to Hong et al~\cite{hong2018you}, repeating patterns are to be found in up to 5.6\% of a typical page execution time.

\idea{Limitations}
This feature category may introduce false positives for JS/WASM heavy sites that perform large amounts of repetitive computation, such as 3d games, 3d ads, and physics simulations, where the scene has to be updated periodically (typically, within the 30-120 times per second range). To discriminate these sites, we have found the final category of features, throttling detection features, to be critical.

\subsubsection{Features: Throttling Detection}
Mining in inherently computationally heavy, and its side effects, such as head and loud fan noise, are easily noticed by cryptojacking victims.
To remain unnoticed for longer, the majority  of miners perform throttling, as we show in Figure~\ref{figure:coinhive_throttling} and Figure~\ref{figure:cpu_usage}.
This reduces the revenue generated by each victim, but this trade-off is compensated by reaching a broader audience.
To perform throttling, miners keep track of the time spent on the generation of each hash and introduce microsleep intervals in their scripts, so to keep the overall execution time under the throttling threshold.

The idea underlying this last set of features is to use throttling, this adaptive behavior that is unique to cryptominers, to detect them.
To craft features that capture throttling we introduce a novel active probing technique: we artificially reduce the amount of CPU time available to the browser, and we load the site suspected of mining. By comparing traces with varying levels of CPU slowdowns, we can identify miners by looking for execution traces where the overall execution time has not changed. This is because legitimate websites naturally will require more execution time to render on a (artificially) less powerful CPU, whereas miners will adapt to the CPU power, and reduce their hash rate to keep their execution time under the throttling threshold. Because throttling is unique to cryptominers, this technique is capable of discriminating miners from legitimate websites with highly-repetitive execution patterns.

\begin{figure}[htbp]
    \centering
    \includegraphics[width=1.0\linewidth]{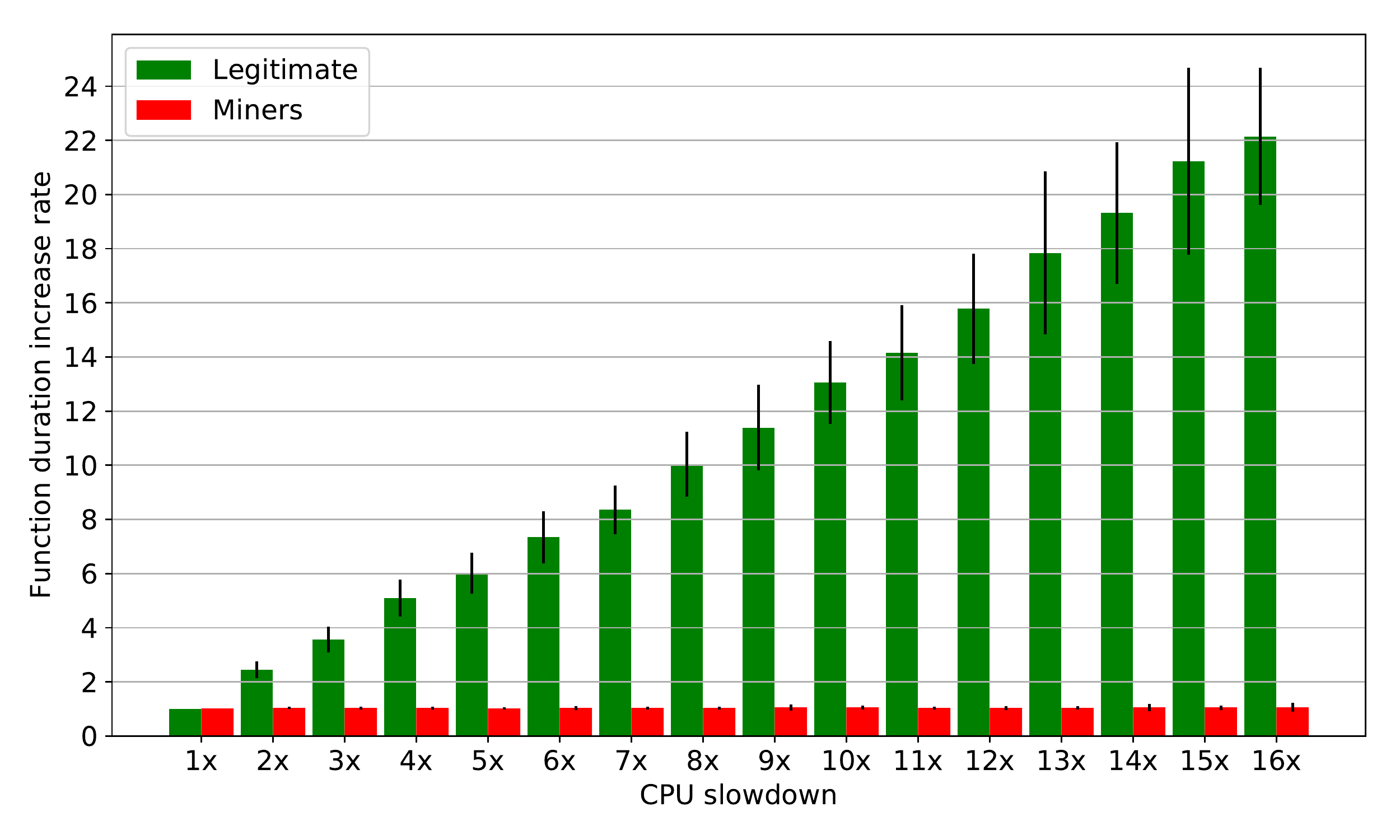}
    \caption{Function duration increase (bars represent mean, whiskers -- standard deviation).}
    \label{figure:durations}
\end{figure}

As a preliminary measurement of this discriminative power, we have manually analyzed traces of publicly available mining scripts, selected video streaming services and JS/WASM games, and we present the results in Figure \ref{figure:durations}, which shows the relation between a CPU slowdown rate and  the average increase of function execution time. Interestingly, the more miners leverage threshold to become unnoticed, the more this behavior becomes evident with this feature. Note that the rate of increase is not exactly linear due to noises introduced by execution randomness, especially in the case of 3d games.

\idea{Limitations}
This category of features is only able to detect throttling. Miners that do not perform throttling, either by mining a set amount of hashes per second or by mining as fast as possible,  will remain unnoticed. However, we argue that unthrottled miners can be easily detected by the previous set of features, and rate-limited miners will have to select a very low hashing rate to remain unnoticed, as this rate is dictated by the slowest executing device (i.e., if the miner aims to only use 20\% of CPU on all devices and does not perform active throttling, the slowest device will determine the hashing rate).

\subsection{Data Collection}
In this section and the next, we will respectively describe how we collect the aforementioned features, and how we encode them in a format that can be digested by our deep-learning classifiers to produce a verdict.

The input of the overall system are URLs of sites suspected of cryptojacking. We load each URL into an instance of the Chrome web browser, which is spawned from scratch for each URL, so to make sure of the independence of each measurement. We control these Chrome instances programmatically through \emph{puppeteer}~\cite{puppeteer}.

Each measurement comprises of four phases. In \ballnumber{1}, we visit the URL and wait until the browser generates less than 4 network requests per second, so to ensure the site is fully loaded. Then (\ballnumber{2}), we wait for ten additional seconds to wait for delayed miners to start working - we chose this particular value because we have found it to be an upper bound of the miner start delay through manual experimentation. We then (\ballnumber{3}) start our measurement phase, which lasts for five seconds. During this time, we collect HPC values through  \emph{perf}~\cite{perf}, and JS/WASM function call log through the  \emph{Chrome DevTools API}~\cite{devtools}.

\idea{HPC traces} We collect the following HPC values, with a sampling interval of 10 milliseconds: CPU cycles, instructions, L1 and L3 cache loads, stores and misses, branch instructions and branch misses. We reset the HPC counters at the beginning of each interval so that the collected values are not cumulative.

\idea{JS/Wasm Traces} We process the DevTools logs to traces where each function is described as a  tuple $\langle$\emph{thread id, function name, source script, line number, column number, call stack depth}$\rangle$. The traces correspond to the JS/WASM function call log in the main execution thread, and any WebWorker threads. Finally (\ballnumber{4}), we produce an event log with the HPC values and JS/WASM function calls (identified with the tuple $\langle$\emph{function name, source script, line number, column number, call stack depth}$\rangle$), to pass on to our feature extractor.

We visit each URL five times, using DevTools to artificially slowing down the CPU available to the site. DevTools emulates slowdowns by introducing short delays during rendering and JS/WASM execution. The slowdown rates we use are 1x (i.e., no slowdown), 2x, 4x, 8x and 16x, which produce a total of five traces per web page.

\subsection{Data Representation}

In this section, we describe how we encode our feature in a format that can be processed by a deep-learning classifier, where each feature has a fixed position in the input array of the classifier.
We note that each classifier will only receive a specific set of features: for example, the classifiers that are based exclusively on HPC values will not be given the JS/Wasm data.

\begin{figure}[htbp]
    \centering
    \includegraphics[width=1.0\linewidth]{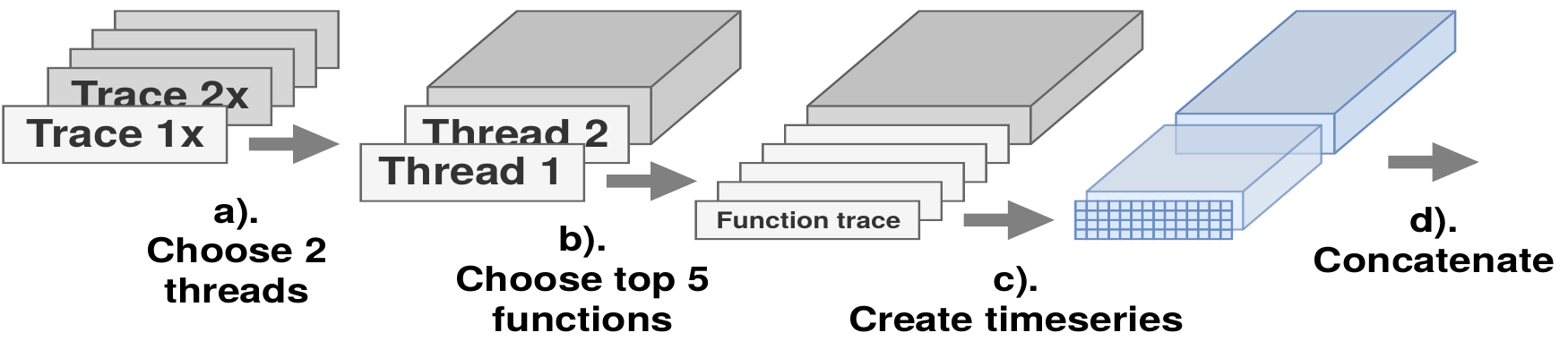}
    \caption{Timeseries creation.}
    \label{figure:timeseries_creation}
\end{figure}

Since the collected HPC values are a fixed set, we represent them as multivariable timeseries where for each counter we have a separate timeseries of its values. Each frame of this timeseries corresponds to a collection interval.
JS/WASM stack traces are also represented as multivariable timeseries, as shown in Figure~\ref{figure:timeseries_creation}.
However, we cannot use the same approach as HPC values, as  each website has its unique set of functions.

We first collect the five traces obtained from visiting a single URL with different CPU slowdown rates (step \textbf{a} in Figure \ref{figure:timeseries_creation}). Each of these traces contains a log of the activity in the main execution thread, plus any spawned WebWorker. If at most one WebWorker has been spawned, we group the main thread trace with the WebWorker (if any), and we proceed to the next step. If more than one WebWorker has been spawned, we create a trace group for each WebWorker, by duplicating the main thread. These trace groups are  passed to the classifier independently: this allows us to use the classifier verdict to pinpoint which WebWorker (or main thread) is likely responsible for the mining activity.

For each trace group, we identify the five functions that have the highest cumulative execution time (step \textbf{b} in Figure \ref{figure:timeseries_creation}), and we discard the rest. We generate a function timeseries for each function, which encapsulates its call times and call durations.

\begin{figure}[htbp]
    \centering
    \includegraphics[width=1.0\linewidth]{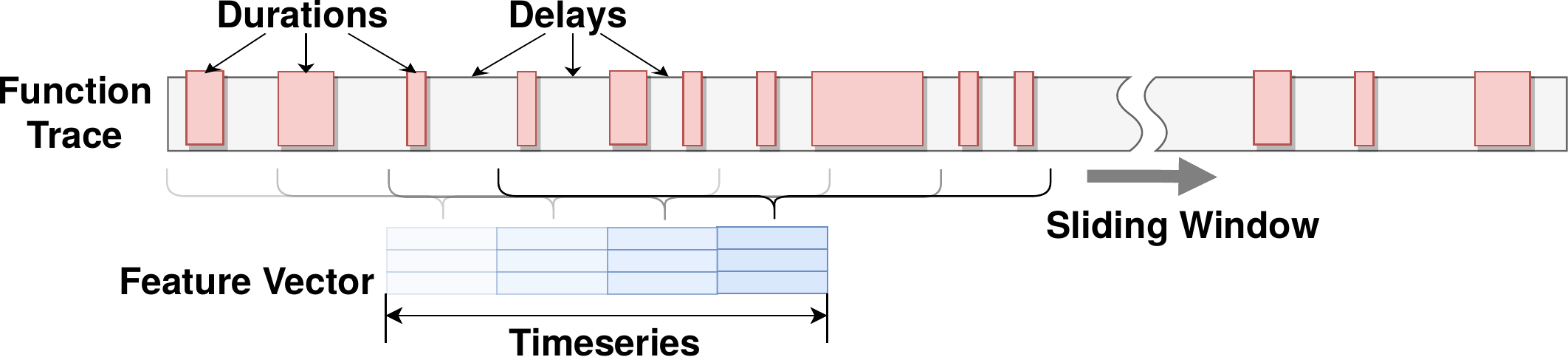}
    \caption{Timeseries for a single function trace.}
    \label{figure:feature_vector}
\end{figure}

Finally, we process these function timeseries  to create a final multivariate timeseries (step \textbf{c} in Figure \ref{figure:timeseries_creation}).
Each timeseries is created using a time-sliding window that traverses a trace one step at a time and creates feature vectors (Figure \ref{figure:feature_vector}).
In  our experiments we use a sliding window size of 2 seconds, and a step of 15 milliseconds.

To better detect the mining of the large family of \emph{SHA-256}-based cryptocurrencies (i.e., not based on CryptoNight), such as \emph{JSEcoin}~\cite{jsecoin}, we also create a separate timeseries describing  calls to the \texttt{WebCrypto} browser module, which is an API to execute selected cryptographic primitives efficiently. Specifically, we log calls to \texttt{DoDigest()}. This timeseries is created using the same sliding-window technique described above.

\subsection{Classifiers}
As you may recall from Section~\ref{section:algos}, we implement a total of seven cryptojacking classifiers. All these classifiers, but the baseline one, are implemented in \emph{Keras}~\cite{keras}, with a \emph{TensorFlow} \cite{tensorflow} backend. We note that all classifiers in related work are not based on neural networks: as we implemented their strategy, we opted to upgrade them to a deep-learning classifier for a fairer comparison.
For the classification task we used two architectures: recurrent networks, and convolutional ones.

\begin{figure}[htbp]
    \centering
    \subfloat[Recurrent.\label{figure:architectures_rnn}]{\centering
        \includegraphics[height=0.5\linewidth]{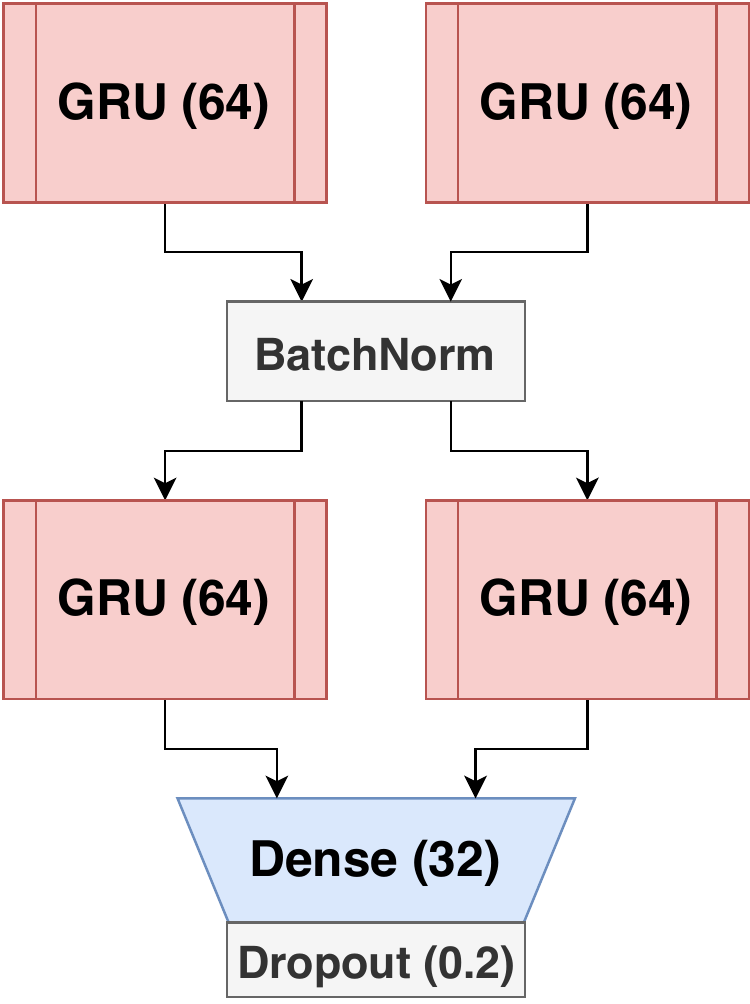}
    }
    \hfill
    \subfloat[Convolutional.\label{figure:architectures_cnn}]{\centering
        \includegraphics[height=0.5\linewidth,width=0.30\linewidth]{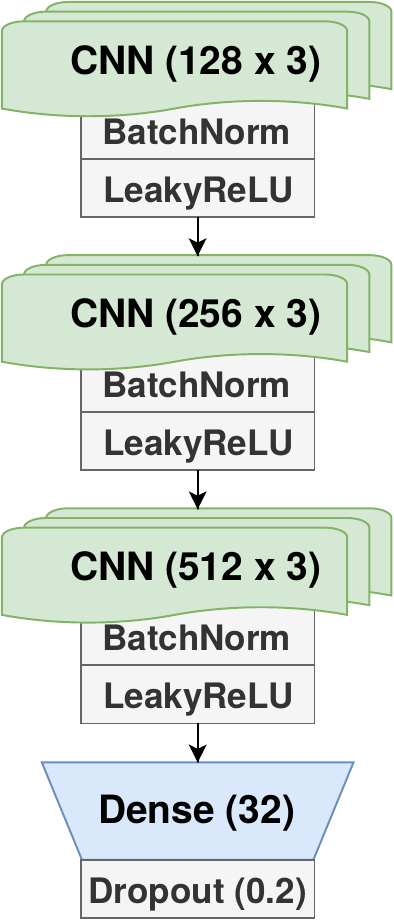}
    }
    \caption{Classifier architectures.}
    \label{figure:architectures}
\end{figure}

We show our recurrent architecture in Figure~\ref{figure:architectures_rnn}). It contains two bidirectional Gated Recurrent Unit layers and employs batch normalization~\cite{ioffe2015batch}. The output from these layers is sent to a final fully-connected layer with 32 neurons and a 20\% dropout~\cite{srivastava2014dropout}.
Such an architecture is commonly in use for text classification tasks.

Our convolutional architecture is shown in Figure~\ref{figure:architectures_cnn}. The various parameters in the figures have been selected through hyperparameter optimization.

\section{Evaluation}\label{section:evaluation}

In this section we evaluate the performance of CoinPolice and perform a large scale investigation on Alexa's top 1 Million websites aimed to find active cryptojacking sites.

\subsection{Dataset}
Our dataset contains 47k samples, and is roughly evenly split between cryptojacking and non-cryptojacking samples.
We collected the legitimate samples at random from the Alexa top million sites, excluding sites that appear in public cryptomining blacklists. To this dataset we added hand-picked video service and t/WASM games, so to introduce hard samples (that is, benign samples that are harder to discriminate from cryptojacking samples).

To create the cryptojacking dataset, we have also manually collected several variants of 23 mining libraries. We then proceeded to visit sites picked from the same Alexa collection mentioned above (but not the same ones), and we performed a local injection of the mining library into the browser, as to emulate a cryptojacking site. This data-augmentation approach lets us generate a large dataset where we control the miner throttling level, which we randomized to diversify the collected data. We added to this dataset a selection of active cryptojacking sites. We collected these by loading sites that were blacklisted for cryptojacking, and manually checking that the mining code was being executed.

To increase the robustness of our classifier,  we have augmented the data in our 47k-samples dataset. Specifically, we augment the data in two ways.
First, we created additional training samples by swapping the main thread with a WebWorker thread. We do so to prevent the classifier from learning that mining typically happens in a WebWorker thread, as attackers could evade detection by moving the computation process across threads.
Second, we created samples with randomly shifted function call times, with shifts up to 25\% of the trace length. We have done so to emulate an evasion technique where the attacker introduces random delays in the execution pattern to evade detection. Some of the related work, such as Outguard, are vulnerable to this attack.
The reason behind leveraging data augmentation is to generate sufficient samples with advanced evasion techniques for our classifiers to learn them.
\subsection{Classifier Evaluation}

Each classifier was trained for 100 epochs by an \emph{Adam} stochastic optimizer \cite{kingma2014adam}, and we performed hyperparameter tuning to obtain the best performance.
After training, we have chosen the best classifiers based on their validation accuracy and evaluated them on our  test data.

Table \ref{table:accuracy} shows training and validation accuracy of best classifiers and their evaluation on test data.
It also shows false negative (\textbf{FNR}) and false positive rates (\textbf{FPR}) evaluated on test data.

\begin{table}[htbp]
  \caption{Accuracy of trained classifiers.}
\begin{center}
  \footnotesize
\begin{tabularx}{\linewidth}{lCCCCC}
  \toprule
    \textbf{Classifier} &
    \textbf{Train acc.} &
    \textbf{Val. acc.} &
    \textbf{Test acc.} &
    \textbf{Test FNR} &
    \textbf{Test FPR} \\
  \midrule
    \textbf{HPC CNN} &
    0.9746 &
    0.8904 &
    0.8865 &
    0.0195 &
    0.2406 \\
    \textbf{HPC RNN} &
    0.9706 &
    0.9701 &
    0.9717 &
    0.0299 &
    0.0261 \\
  \midrule
    \textbf{JS CNN (No Slowdown)} &
    0.9097 &
    0.8361 &
    0.8441 &
    0.0799 &
    0.2478 \\
    \textbf{JS RNN (No Slowdown)} &
    0.8637 &
    0.8705 &
    0.8648 &
    0.0545 &
    0.2355 \\
  \midrule
    \textbf{JS CNN} &
    0.9873 &
    0.9783 &
    0.9734 &
    0.0379 &
    0.0131 \\
    \textbf{JS RNN (CoinPolice)} &
    0.9830 &
    0.9779 &
    0.9781 &
    0.0353 &
    0.0074 \\
  \bottomrule
\end{tabularx}
\label{table:accuracy}
\end{center}
\end{table}

The results show that convolutional classifier based on HPC shows an unacceptable amount of overfitting, while a recurrent classifier shows high test accuracy rate.
But it still shows visible signs of overfitting such as validation accuracy oscillations with a high amplitude (accuracy could drop below 93\%).

The results also show the importance of an artificial CPU slowdown for detecting miners that throttle.
Without features corresponding to such slowdowns test accuracy drops below 87\% due to high false positive rates (more than 23\%).

Both JS based classifiers (with CPU slowdowns) show higher values of each accuracy type.
JS CNN starts overfitting (validation accuracy starts to decrease) after 40 epochs.
JS RNN classifier is more stable than JS CNN and shows less overfitting by checking on validation data.

To evaluate the ability to detect throttled miners we have checked true positives on a separate test set consisting of $\sim$22k mining samples with different throttling rates.
We have also evaluated an existing threshold-based classifier presented in \cite{hong2018you} that detects functions that use more than 30\% of the CPU time.
It's important to note that this data set was collected after the Monero hard fork happened at the 9\textsuperscript{th} of March \cite{monerolatest}.
Thus, it includes fewer miner types since several miners, including Coinhive, have stopped working \cite{coinhiveclosed}.

\begin{figure}[thbp]
    \centering
    \includegraphics[width=1.0\linewidth]{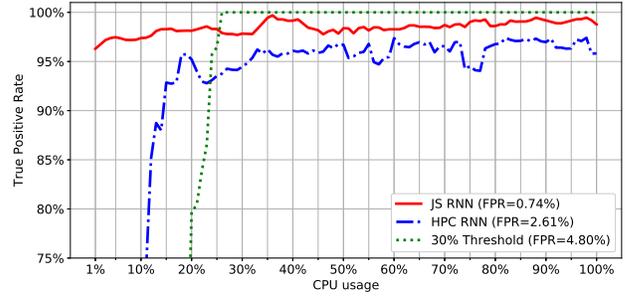}
    \caption{Detection of throttled miners.}
    \label{figure:throttling}
\end{figure}

The results are shown in Figure \ref{figure:throttling}.
This figure shows true positive rates for JS RNN and HPC RNN  classifiers and for the 30\% threshold-based classifier.

Experiments have shown that the JS-based classifier is able to detect silent miners ($\leqslant30\%$ of the CPU) with the smallest amount of false positives ($0.74\%$).

Even though the threshold-based classifier shows $100\%$ true positive rate on heavy miners, it starts to drop instantly on miners that use $\leqslant25\%$ of the CPU.
Results also show that the classifier based on hardware performance counters is not able to correctly classify miners that use $\leqslant15\%$ of the CPU.

\subsection{Investigation}

We have performed a large scale investigation on Alexa's top 1 Million websites.
We have opened each website and have collected traces of JS and WASM calls using Chrome DevTools API \cite{devtools} and the \emph{puppeteer} framework \cite{puppeteer}.

Data was collected using 11 servers with $\sim$16 CPUs each.
We have collected $\sim$2.2 Tb of zipped data during the period between 29\textsuperscript{th} of December 2018 and 23\textsuperscript{rd} of January 2019.

It should be noted that in order to increase sample collection speed we spoofed the \texttt{navigator hardwareConcurrency} object to return only 1 CPU thread available.
This object allows JS to get the number of logical processors.
Such spoofing allowed us to profile multiple web pages in parallel.
So if we would profile several miners at the same time they would use only a single thread and would not consume all threads available simultaneously.

To reduce the number of potential false positives during the investigation, we didn't use model on samples that were using almost 0\% of the CPU, since they are obviously not miners.
Also we didn't click any buttons on web pages since we only searched for malicious mining without a user permission.

The results of our investigation show that 447 websites from the Alexa's top 1 Million list are mining cryptocurrency, which we manually confirmed.

\begin{figure}[htbp]
    \centering
    \includegraphics[width=1.0\linewidth]{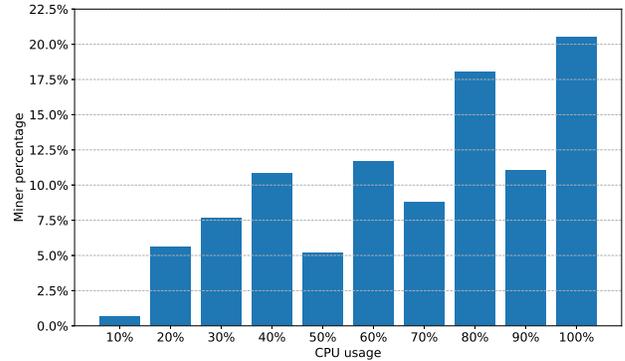}
    \caption{CPU usage by mining scripts.}
    \label{figure:cpu_usage}
\end{figure}

Our approach also allows to show how much real CPU time is used for mining process since data contains information about function durations.
Figure \ref{figure:cpu_usage} shows the results of our investigation.
The results show that even though $50\%$ of all miners in Alexa's list are using $\geqslant80\%$ of the CPU there are miners that are very silent ($\leqslant30\%$ of the CPU).

We have also checked CoinPolice on URLs that contained code patterns corresponding to well-known mining scripts.
We have collected these URLs with pattern matching using \emph{PublicWWW}~\cite{publicwww} on 31\textsuperscript{st} of January 2019.
The Table~\ref{table:patterns} shows patterns that were used in the search.
It's important to note that we only considered web pages that could be successfully opened.

Tables \ref{table:publicwww_alexa} and \ref{table:publicwww} show the number of pages for each mining script and also show the results of CoinPolice classification.
It's important to note that Table \ref{table:publicwww_alexa} shows the subset of websites presented in Table \ref{table:publicwww} that are in the Alexa Top 1M list.

\begin{table}[htbp]

  \caption{Miner detection on PublicWWW data (Top 1M).}
\begin{center}
  \footnotesize
\begin{tabularx}{\linewidth}{lRRR}
  \toprule
    \textbf{Miner} &
    \textbf{Top 1M} &
    \textbf{Actively mining} &
    \textbf{Active \%} \\
  \midrule
    \textbf{Coinhive} & 159 & 92 & 57.9\% \\
    \textbf{JSEcoin} & 103 & 4 & 3.9\% \\
    \textbf{Crypto-Loot} & 43 & 18 & 41.9\% \\
    \textbf{Mineralt} & 21 & 8 & 38.1\% \\
    \textbf{CoinImp} & 20 & 15 & 75\% \\
    \textbf{OMINE} & 13 & 10 & 76.9\% \\
    \textbf{Perfekt} & 12 & 8 & 66.7\% \\
    \textbf{WebMine} & 4 & 3 & 75\% \\
    \textbf{WebMinePool} & 3 & 1 & 33.3\% \\
    \textbf{CoinNebula} & 2 & 2 & 100\% \\
    \textbf{Minero} & 2 & 2 & 100\% \\
  \midrule
    \textbf{Total} & 382 & 163 & 42.7\% \\
  \midrule
    \textbf{Unique} & 376 & 157 & 41.8\% \\
  \bottomrule
\end{tabularx}
\label{table:publicwww_alexa}
\end{center}
\end{table}

\begin{table}[htbp]
\caption{Miner detection on PublicWWW data}
\begin{center}
  \footnotesize
\begin{tabularx}{\linewidth}{lRRR}
  \toprule
    \textbf{Miner} &
    \textbf{All samples} &
    \textbf{Mining} &
    \textbf{Percentage} \\
  \midrule
    \textbf{Coinhive} & 10771 & 5222 & 48.5\% \\
    \textbf{JSEcoin} & 1939 & 131 & 6.8\% \\
    \textbf{CoinImp} & 785 & 194 & 24.7\% \\
    \textbf{Crypto-Loot} & 733 & 441 & 60.2\% \\
    \textbf{OMINE} & 305 & 192 & 63\% \\
    \textbf{Mineralt} & 230 & 96 & 41.7\% \\
    \textbf{Perfekt} & 151 & 119 & 78.8\% \\
    \textbf{Minero} & 121 & 102 & 84.3\% \\
    \textbf{WebMinePool} & 69 & 40 & 58\% \\
    \textbf{CoinNebula} & 61 & 39 & 63.9\% \\
    \textbf{WebMine} & 37 & 20 & 54.1\% \\
  \midrule
    \textbf{Total} & 15202 & 6596 & 43.4\% \\
  \midrule
    \textbf{Unique} & 14891 & 6410 & 43\% \\
  \bottomrule
\end{tabularx}
\label{table:publicwww}
\end{center}
\end{table}

During this investigation we have found 6410 mining websites 157 of which are in the Alexa's top 1 Million list.
Table \ref{table:publicwww} also shows that 186 websites are using several miner types simultaneously.

The results show that only 43\% of websites, that contain string patterns corresponding to miners, are actually mining cryptocurrency.
There are several possible explanations behind this.

First, as we found by a manual check, a lot of websites do not run miners because mining script could not be loaded successfully.
This could be because proxies for servers serving mining scripts are changed in a constant manner to evade being blocked by ISPs or by mining blocking software on the user side.
So hosts of mining websites need to constantly update code for miner loading.

Second, some scripts could have been inserted only once though a vulnerability on a website and haven't been updated after this.

Third, some websites are asking for user permission to start mining, so they do contain mining code but they do not run it right away.
For example, this is the case for JSEcoin, since it requires a manual change in the official script to disable user permission check.
So only 6.8\% of them are mining maliciously.

Forth, Monero cryptocurrency network introduced a hard-fork and updated CryptoNight algorithm in October 2018 \cite{moneronew}.
These could become a problem for someone who was using self-hosted mining pools on personal servers to avoid using well-known proxies or to avoid commission from official pools (i.e. 30\% from Coinhive).
So all of them needed to update mining scripts, since old scripts got deprecated.

Combining these 2 investigations together we have found 6700 mining websites 447 of which are in the Alexa's top 1 Million list.
We also have checked our system on data from previous large scale investigation shown in \cite{hong2018you}.
Their dataset consisted of 2770 mining websites (2580 unique domains) and was collected at the beginning of the 2018.
Our experiments have shown that only 381 of them are still mining cryptocurrency (113 of which are in the Alexa's top 1 Million list in January 2019).

To quantify the impact of the Coinhive's discontinuation \cite{coinhiveclosed} and to capture how cryptomining sites react to big changes in the environment, we have re-scanned all Coinhive URLs that were originally found by CoinPolice.
The experiment took place 3 weeks after the discontinuation, so websites' owners would have enough time to replace Coinhive with another working miner.

The results are shown in Table \ref{table:coinhive}.
Experiments have shown that more than 98\% of mining URLs have stopped mining, i.e CoinHive has stopped functioning and has not been replaced with a new miner, making the recovery rate of cryptojacking websites is $\sim$1.2\%.
The are multiple possible explanations for this low recovery rate: the initial miner might have been surreptitiously placed on the site through a vulnerability, the owner did not notice that CoinHive ceased operations, or the profit obtained from cryptojacking is not notable enough to warrant replacing the mining logic.

\begin{table}[thbp]
\caption{CoinHive's discontinuation impact}
\begin{center}
  \footnotesize
\begin{tabularx}{\linewidth}{lRRRR}
  \toprule
    \textbf{Date} &
    \textbf{URLs} &
    \textbf{Reachable} &
    \textbf{Mining} &
    \textbf{Recovery} \\
  \midrule
    \textbf{28\textsuperscript{th} of March} & 5222 & 4863 & 66 & 1.26\% \\
    \textbf{11\textsuperscript{th} of April} & 5222 & 4859 & 61 & 1.17\% \\
    \textbf{18\textsuperscript{th} of April} & 5222 & 4845 & 63 & 1.21\% \\
  \bottomrule
\end{tabularx}
\label{table:coinhive}
\end{center}
\end{table}

\section{Discussion}

In this section we discuss limitations of our approach and potential evasion techniques that can be used by miners to defeat detection. We identify several limitations in the way we collect data. These limitations are the result of a trade-off of the resources we spend on a single site and the number of sites processed. As such, they can be mitigated.

\idea{Timing} We only profile each website for $5$ seconds ($+10$ second delay before profiling) we could possibly miss cryptojacking websites that start mining after more than $15$ seconds. The cryptojacker faces a trade-off too here, as this delay will indeed reduce the chance of detection but it will impact the profits generated.

\idea{Profiling} As we use the Chrome DevTools API we could miss websites that detect the usage of DevTools. This can be mitigated by patching Chrome to evade these detections.

\idea{Functions in timeseries} We only include the top five functions in terms of CPU usage per thread in our timeseries.
So if a website’s owner implements additional logic in the miner’s thread, and this logic is computationally heavier than the miner, then we could miss mining behavior. We manually checked the top mining families for this behavior, and we found that none of them contains it. Also, the number of functions included in the timeseries is configurable.

We envision several ways in which CoinPolice could be evaded.

\idea{Per-request randomization of JS/WASM code}
A  web page may randomize all function names at each page load, so we will not be able to compare function durations in traces corresponding to different CPU throttling rates. This behavior, if adopted by cryptojackers, might become a feature in itself. If so, CoinPolice could be reformulated to identify recurring code execution without using the function names, such as by convolving the flame graphs.

\idea{Randomized function delays}
Miners may introduce random delays between function calls to evade the throttling-detection features.
We have considered this attack in our training, test and validation data that we presented: through data augmentation, we randomly shifted the execution of function calls. However, we were unable to check our performance on real data, as none of the existing miners implement this technique.

\idea{Randomized function durations}
Miners also can change function durations depending on the CPU performance, so these durations would not be the same in the presence of an artificial CPU slowdown. This technique would effectively evade the throttling-detection features, yet we argue it would ultimately be detrimental to the cryptojacker. This is because on sufficiently slow devices the mining speed will have to use 100\% of the available CPU, thus becoming very noticeable.
 \section{Conclusion}

In this paper we present CoinPolice -- a system for hidden cryptojacking detection.
We have identified features that can be used to detect malicious cryptocurrency mining on web pages.
We also developed a new active probing technique that can detect hidden miners that are trying to evade detection by reducing their CPU usage.

Based on the identified features and a deep neural network classifier we have developed a system called CoinPolice that can detect 97.87\% of hidden miners with a low false positive rate (0.74\%).
A performance comparison against the current state of the art has shown that CoinPolice outperforms it in detecting highly throttled miners with lower amount of false alarms.

We have evaluated CoinPolice on existing miners and have performed a large scale investigation and have found 6700 mining websites 447 of which are in the Alexa’s Top 1 Million list.
We have also checked mining websites found by pattern matching in the PublicWWW base.
Our results show that only 43\% of these websites are successfully mining cryptocurrency.

\begin{table}[htbp]
\caption{Miner code patterns}
\label{table:patterns}
\begin{center}
  \footnotesize
\begin{tabularx}{\linewidth}{lX}
  \toprule
    \textbf{Miner} &
    \textbf{Patterns} \\
  \midrule
    \textbf{Coinhive} &
    \makecell[tl]{"//coinhive.com/lib", "//coin-hive.com/lib",\\ "//coinhive.com/captcha", "//coin-hive.com/captcha",\\ ".coinhive.com/proxy", ".coin-hive.com/proxy",\\ "CoinHive.Anonymous", "CoinHive.User"} \\
  \midrule
    \textbf{JSEcoin} &
    \makecell[tl]{"//jsecoin.com/server", "//server.jsecoin.com/",\\ "//load.jsecoin.com/", ".jsecoin.com/server",\\ ".server.jsecoin.com/", ".load.jsecoin.com/",\\ "jsecoin.com/load", "JSE.User"} \\
  \midrule
  \textbf{CoinImp} & \makecell[tl]{"www.coinimp.com/scripts/min.js", "www.hashing.win",\\ "www.hostingcloud.racing", "coinimp.Anonymous"} \\
  \midrule
    \textbf{Crypto-Loot} &
    \makecell[tl]{"//crypto-loot.com/lib", ".crypto-loot.com/proxy",\\ ".webmine.pro/", "//cryptoloot.pro/lib/",\\ "//cryptaloot.pro/lib/crypta.js", "CryptoLoot.Anonymous",\\ "CryptoLoot.User", "CRLT.Anonymous", "CRLT.User"} \\
  \midrule
    \textbf{OMINE} &
    \makecell[tl]{"throttleMiner", "OMINE(", "OMINEId(",\\ "xmr.omine.org/assets/"} \\
  \midrule
    \textbf{Mineralt} &
    \makecell[tl]{"mepirtedic.com/amo.js", "play.besstahete.info/app.js",\\ "https://play.feesocrald.com/app.js", "ecart.html?bdata"} \\
  \midrule
    \textbf{Perfekt} &
    \makecell[tl]{"perfekt/perfekt.js", "PerfektStart", "EverythingIsBinary"} \\
  \midrule
    \textbf{Minero} &
    \makecell[tl]{"minero.cc/lib/", "minero.pw/miner.min.js",\\ "minero.Anonymous"} \\
  \midrule
    \textbf{WebMinePool} &
    \makecell[tl]{"webminepool.com/lib/base.js",\\ "webminepool.com/lib/captcha.js",\\ "webminepool.com/api/", "webminepool.tk/",\\ "WMP.Anonymous", "WMP.User"} \\
  \midrule
    \textbf{CoinNebula} &
    \makecell[tl]{"coinnebula.com/lib/", "1q2w3.website/lib/",\\ "CoinNebula.Instance"} \\
  \midrule
    \textbf{WebMine} &
    \makecell[tl]{"webmine.cz/miner", "webmine.cz/worker",\\ "authedwebmine.cz/authedminer.js"} \\
  \bottomrule
\end{tabularx}

\end{center}
\end{table}
 \balance
\bibliographystyle{ACM-Reference-Format}
\bibliography{bibliography}
 \end{document}